\documentclass[10pt]{article} 

\usepackage[preprint]{rlj} 

\usepackage{amssymb}            
\usepackage{mathtools}          
\usepackage{mathrsfs}           
\usepackage{graphicx}           
\usepackage{subcaption}         
\usepackage[space]{grffile}     
\usepackage{url}                
\usepackage{lipsum}             

\usepackage{amsthm}
\usepackage{multirow}

\usepackage{cite}
\usepackage{amsmath,amssymb,amsfonts}
\usepackage{graphicx}
\usepackage{textcomp}
\usepackage{xcolor}
\def\BibTeX{{\rm B\kern-.05em{\sc i\kern-.025em b}\kern-.08em
    T\kern-.1667em\lower.7ex\hbox{E}\kern-.125emX}}

\usepackage{algpseudocode}
\usepackage{float}
\usepackage{soul}
\usepackage{url}
\usepackage{color}
\usepackage{courier}
\usepackage{multicol}
\usepackage[utf8]{inputenc}
\usepackage{graphicx}
\usepackage{amsmath}
\usepackage{amsthm}
\usepackage{amssymb}
\usepackage{bbm, bm}
\usepackage{booktabs}
\usepackage{setspace}
\usepackage{adjustbox}
\usepackage{algorithm}

\newcommand{\alert}[1]{\textbf{}}

\newcommand{\BEAS}{\begin{eqnarray*}}
\newcommand{\EEAS}{\end{eqnarray*}}
\newcommand{\BEA}{\begin{eqnarray}}
\newcommand{\EEA}{\end{eqnarray}}
\newcommand{\BEQ}{\begin{equation}}
\newcommand{\EEQ}{\end{equation}}
\newcommand{\BIT}{\begin{itemize}}
\newcommand{\EIT}{\end{itemize}}
\newcommand{\BNUM}{\begin{enumerate}}
\newcommand{\ENUM}{\end{enumerate}}
\newcommand{\BEL}[1]{\begin{equation}\label{#1}}
\newcommand{\EEL}{\end{equation}}

\newcommand{\state}{\mathbf{s}}

\newcommand{\action}{\mathbf{a}}

\newcommand{\policy}{\pi}

\newcommand{\reward}{r}
\newcommand{\rewmodel}{\widehat{\reward}_\psi}

\newcommand{\BA}{\begin{array}}
\newcommand{\EA}{\end{array}}

\DeclareMathOperator*{\expec}{\mathbb{E}}

\DeclareMathOperator{\sign}{sign}

\usepackage{xspace}
\newcommand{\alggen}{Crowd-PrefRL\xspace}
\newcommand{\alg}{Crowd-PrefPPO\xspace}
\newcommand{\newsec}[1]{\vspace{2mm} \noindent \textbf{#1.} }

\theoremstyle{definition}

\theoremstyle{remark}

\title{\alggen: Preference-Based Reward Learning from Crowds}

\setrunningtitle{\alggen: Preference-Based Reward Learning from Crowds}

\author{David Chhan\textsuperscript{1,$\dagger$}, Ellen Novoseller\textsuperscript{1}, Vernon J. Lawhern\textsuperscript{1}}

\emails{david.chhan.civ@army.mil}

\affiliations{
$^{1}$\textbf{DEVCOM Army Research Laboratory}
}

\contribution{
    We propose a conceptual framework for preference-based RL from crowds,
    Crowd-PrefRL, that integrates preference-based reward learning approaches with techniques derived from unsupervised crowdsourcing to enable the training of autonomous agents with feedback from crowds of unknown expertise and reliability.
    }
    {
    Prior works on RL from human feedback (RLHF), e.g.~\citet{Ouyang2022}, typically aggregate preference data across users without collecting multiple preference labels for each data point and without recording the associated annotator identifiers. In contrast, this work studies the aggregation of preference labels under the assumptions that multiple users label each user query and that annotator identifiers are captured. This enables inference of a noise-robust, aggregated crowd preference label for each query, i.e., predicting the ``true'' preference labels with respect to an inferred underlying reward function, while also detecting the presence of minority feedback groups (who provide feedback according to a different reward function) using techniques derived from unsupervised crowdsourcing.
    }

\contribution{
    We demonstrate empirically that Crowd-PrefRL is capable of (1) learning reward functions from crowd feedback that are more accurate than reward functions learned from any individual member of the crowd, (2) learning reward functions that are more accurate than reward functions learned from a majority vote, and (3) identifying \textit{minority feedback groups} within the crowd in an unsupervised manner, given only the crowd preference data, for instance to cluster users according to their objectives. 
    }
    {
    Prior methods generally make certain modeling assumptions (e.g. the number of distinct minority groups, access to offline data), while our method identifies
    minority groups with online RL feedback with small batches of labeled data. 
    }


\summary{Preference-based reinforcement learning (RL) provides a framework to train autonomous agents using human feedback through preferences over pairs of behaviors,
enabling agents to learn desired behaviors when it is difficult to specify a numerical reward function.
While this paradigm leverages human feedback,
it typically 
treats the feedback as given by a single human user. However, different users may desire multiple agent behaviors and modes of interaction.
Meanwhile, incorporating preference feedback from crowds (i.e. ensembles of users) in a robust manner remains a challenge, and the problem of training RL agents using feedback from multiple human users remains understudied. In this work we propose a conceptual framework for preference-based RL from crowds, Crowd-PrefRL, that integrates preference-based reward learning approaches with techniques derived from
unsupervised crowdsourcing to enable the training of autonomous agents with feedback from crowds of unknown expertise and reliability. We show that in most cases, agents trained with \alggen 
outperform agents trained with majority-vote preferences or preferences from any individual user, especially when the spread of user error rates among the crowd is large. 
Results further suggest that our method can identify the presence of minority viewpoints within the crowd in an unsupervised manner.

}

\begin{document}

\maketitle  

\begin{abstract}
Preference-based reinforcement learning (RL) provides a framework to train AI agents using human feedback through preferences over pairs of behaviors, enabling agents to
learn desired behaviors when it is difficult to specify a numerical reward function. While
this paradigm leverages human feedback, it typically treats the feedback as given by a
single human user. However, different users may desire multiple AI behaviors and
modes of interaction. Meanwhile, incorporating preference feedback from crowds (i.e.
ensembles of users) in a robust manner remains a challenge, and the problem of training RL agents using feedback from multiple human users remains understudied. In this work, we introduce a conceptual framework, Crowd-PrefRL, that integrates preference-based RL approaches with techniques from unsupervised crowdsourcing to enable training of autonomous system behaviors from crowdsourced feedback. We show preliminary results suggesting that Crowd-PrefRL can learn reward functions and agent policies from preference feedback provided by crowds of unknown expertise and reliability. We also show that in most cases, agents trained with Crowd-PrefRL outperform agents trained with majority-vote preferences or preferences from any individual user, especially when the spread of user error rates among the crowd is large. Results further suggest that our method can identify the presence of minority viewpoints within the crowd in an unsupervised manner.
\end{abstract}

\section{Introduction}
\label{sec:introduction}
Reinforcement learning from human feedback (RLHF) \citep{casper2023open} is a promising approach for learning intelligent behaviors in the absence of a known numerical reward signal, where humans typically answer pairwise preference queries of the form ``Do you prefer A or B?''~\citep{christiano2017deep,lee2021pebble}. RLHF methods have shown success in domains from robotics~\citep{wilde2021learning,torne2023breadcrumbs,lee2021bpref} to training large language models (LLMs)~\citep{Ouyang2022}. 
Yet, most current RLHF methods treat the preference feedback as if it came from a single human user, even if it is actually given by multiple users with potentially different backgrounds and levels of expertise. In contrast, aggregating crowdsourced data in a manner that recognizes differences between individuals could enable algorithms to analyze disagreement patterns across the crowd, detect minority groups, mitigate model bias and ensure fairness~\citep{Ouyang2022}, and assist users with diverse preferences or varying degrees of noise in their feedback, whether through personalization or by learning behaviors that balance various users' needs.
Indeed, aggregating information across a crowd can result in better decisions than those of any single crowd member \citep{Surowiecki2005}; this has been demonstrated across many domains \citep{Prelec2017Nature} as the ``wisdom of crowds'' phenomenon.

We consider RL from human pairwise preference feedback that is \textit{crowdsourced}, i.e., where (1) each data point is labeled by multiple human users  and (2) these labels are aggregated to form an ensemble label (or, if applicable, identification of minority feedback groups with their ensemble labels) for downstream training. 
Notably, several recent RLHF works propose methods to learn from multi-user feedback, e.g.~\citet{siththaranjan2023distributional, poddar2024personalizing, chakraborty2024maxminrlhf}. However, these prior approaches do not leverage crowdsourcing methods \citep{Chen2013, Li2016} that explicitly model information from different users by incorporating annotator identifiers in the learning process; such methods could potentially augment
RLHF methods with interesting properties, for example characterizing user accuracies, improving data requirements with respect to identifying multiple viewpoints, efficiently assigning users to inferred clusters based on viewpoint, and tailoring queries to specific users. 
In addition, while there are datasets available that study diverse preferences in multi-human feedback \citep{bai2022traininghelpfulharmlessassistant, durmus2024measuringrepresentationsubjectiveglobal, nakano2022webgptbrowserassistedquestionansweringhuman}, these typically do not contain annotator identifiers together with data labeled by multiple annotators, for example due to cost reasons \citep{Ouyang2022}, potentially inhibiting the ability to learn and incorporate diverse preferences among human users. 

\begin{figure}[t!]
\centering
\includegraphics[width=8.5cm]{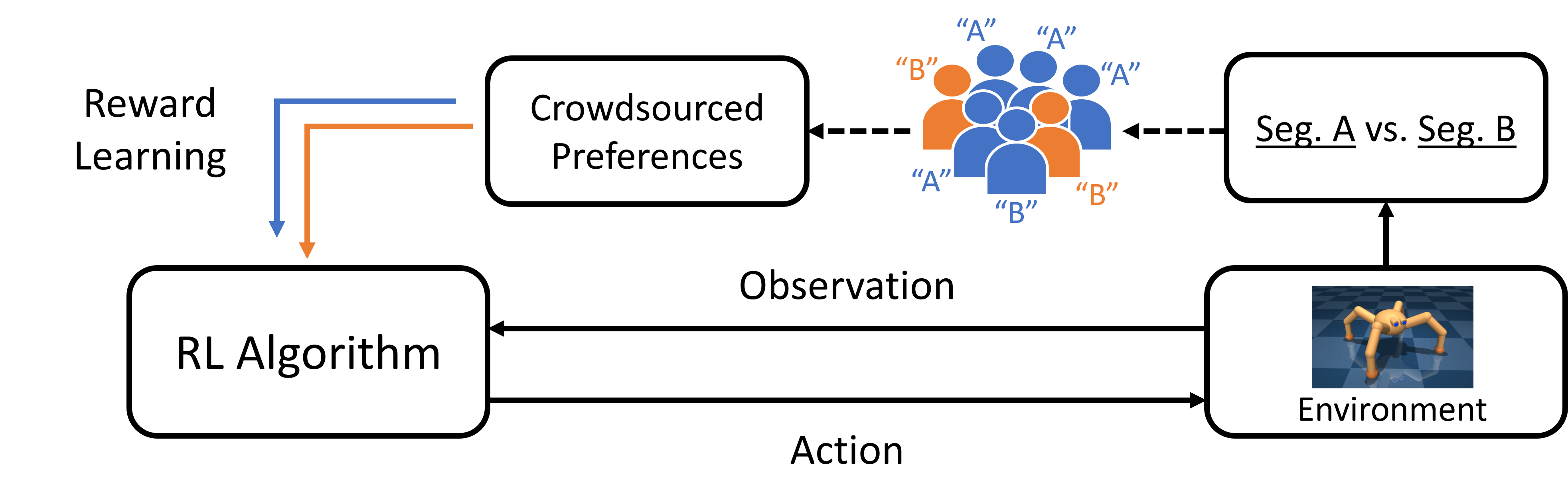}
\vskip -0.1 true in
\caption{\alggen framework for training RL agents 
via crowd feedback. We assume a \textit{crowd}, or ensemble, of users is queried for their preferences over pairs of segmented behaviors (A vs. B). These crowd preferences are then used to learn aggregate preference labels and reward function(s) for preference-based RL. Note that crowds might consist of users who provide diverse preference feedback according to different objectives (\textcolor{blue}{blue} vs. \textcolor{orange}{orange}), which we assume are unknown \textit{a priori}.}
\label{fig:PbRL}
\end{figure}

This work studies preference-based RL leveraging crowd feedback in the \textit{online RL} scenario, where the RL algorithm has environment access and can sample new preference pairs for the crowd to label throughout model training. Unlike prior methods, we propose to perform label aggregation via the Spectral Meta-Learner (SML)~\citep{parisiSML} to learn noise-robust aggregate preference labels to train an aggregate reward. This approach, which leverages annotator identifiers, requires minimal modeling assumptions, is straightforward to compute, and has no hyperparameters to tune. Furthermore, it is robust to multiple viewpoints and as a byproduct, yields weights that enable principled inference of user accuracies and direct clustering of users via their inferred objectives. 
Specifically:
\begin{enumerate}
    \item We introduce a conceptual RL framework, \alggen, that integrates online preference-based RL with techniques from unsupervised crowdsourcing to train AI agents with feedback from crowds of unknown expertise and reliability. \alggen aggregates the user feedback to learn crowdsourced preference labels, and then uses those labels for reward function learning in preference-based RL (see Figure \ref{fig:PbRL}). \alggen can also detect the presence of multiple, e.g. minority, viewpoints among the crowd, given only the crowd's binary preference feedback.
    \item 
    We perform exploratory experiments in robot locomotion tasks suggesting that 
    RL agents trained with crowd preferences outperform RL agents trained via majority-vote preferences or preferences from any individual crowd member, in particular when the spread of user errors among the crowd is large. We also show a proof-of-concept for the ability of our method to identify minority viewpoints and cluster users by viewpoint, enabling downstream personalization to diverse users. 
\end{enumerate}

\section{Related Work}

\newsec{RLHF} While RL from pairwise preference queries has demonstrated numerous successes~\citep{christiano2017deep, lee2021pebble, lee2021bpref, Ouyang2022}, much prior work assumes all data comes from a single human user, in some cases treating multi-user data as if it came from just one person, e.g. as in \citet{Ouyang2022}. In addition, existing publicly available datasets for studying diverse human preferences omit information about annotator identifiers, e.g. \citet{bai2022traininghelpfulharmlessassistant, nakano2022webgptbrowserassistedquestionansweringhuman, durmus2024measuringrepresentationsubjectiveglobal}.
Unlike these works, our approach explicitly models feedback from multiple users with different degrees of noise and who potentially give feedback with respect to multiple reward functions. Our label aggregation approach could be paired together with any RLHF method, including methods that do not learn a reward function~\citep{rafailov2023direct}.

\newsec{Label Aggregation} Previous crowdsourcing literature in classification settings aggregates labels across users to learn crowd labels and detect patterns of outlier feedback ~\citep{Chen2013,Li2016}. \citet{parisiSML} develop the SML label aggregation method, which is robust to both noisy users and to the presence of malicious user groups.
Yet, the application of label aggregation to both preference-based learning and to RL remains understudied.
While \citet{brown2019extrapolating} aggregate labels via a majority vote in an imitation learning setting, unlike the SML, a majority vote cannot account for patterns across users, e.g. users who consistently provide poor feedback. In contrast, the SML performs a spectral analysis of the covariance matrix of the crowd labels to estimate crowd worker reliabilities and can identify the presence of minority user groups in the crowd.

\newsec{RL from Multi-Human Feedback}
Several recent works~\citep{Zhang2023, siththaranjan2023distributional, poddar2024personalizing} use data collected from multiple humans for preference-based reward learning. However, these works are limited to the offline RL setting, which potentially limits the expressiveness of the learned reward to behaviors in the offline dataset.
Furthermore, \citet{siththaranjan2023distributional} and \citet{poddar2024personalizing} infer a distribution over learned rewards across users, potentially requiring restrictive modeling assumptions and large quantities of data. E.g.,~\citet{siththaranjan2023distributional} propose two choices of reward distribution, while ~\citet{poddar2024personalizing} leverage variational inference to infer the reward distribution, incurring higher data requirements compared to the SML (7,500 preference pairs per user). In contrast, our method performs on-policy RLHF and learns aggregate preference labels to train an aggregate reward function. We utilize the SML as a label aggregation method, leveraging annotator identifiers unlike prior RLHF works. In addition, our method provides a principled way to infer user clusters and to infer each user's accuracy with respect to either the majority reward or a particular user cluster's reward. Furthermore, the SML is straightforward to compute, does not have any hyperparameters to tune, and is robust to the presence of multiple groups of users with differing objectives. 

Several of these works pursue pluralistic preference alignment~\citep{poddar2024personalizing, chakraborty2024maxminrlhf, ramesh2024group}, i.e., to train AI models such as LLMs to generate responses that balance a diversity of human viewpoints. Yet, these methods often require modeling assumptions such as a specified number of user clusters or set of user beliefs~\citep{chakraborty2024maxminrlhf, yao2025no, ramesh2024group, zhao2024group}. Also, they typically do not provide flexibility in how to balance the multiple user objectives, e.g. via allowing multiple objective weightings. For instance,~\citet{chakraborty2024maxminrlhf, ramesh2024group} attempt to maximize the minimum utility over all individuals.
Unlike these works, our approach enables user cluster inference, which could facilitate a range of downstream methods for balancing among the learned rewards.

\section{Problem Formulation}
In this paper, we are concerned with answering the following questions: (1) How can we effectively combine preference feedback from multiple users of unknown expertise and reliability in the absence of ground-truth information? (2) Can we learn a crowdsourced reward function that yields better performance
than any individual user's reward function?, and (3) Can we detect the presence of minority feedback (as opposed to purely noisy feedback) in the crowd and cluster users by their objectives using only the preference feedback? This section formulates the problem more precisely. 

We consider an RL agent that takes actions and interacts with an environment \citep{sutton2018reinforcement}, such that at each timestep $t$, the agent receives a state $\state_t$ from the environment and chooses an action $\action_t$ based on its policy $\policy$. In traditional episodic RL, the environment also returns a reward $\reward (\state_t, \action_t)$, and the agent seeks to maximize the total discounted sum of rewards over an episode. 
However, for many complex tasks, it is difficult to construct a suitable reward function. Preference-based RL approaches instead query a human user to obtain preference feedback over segments of agent behaviors and use this feedback to learn a reward function  \citep{christiano2017deep,lee2021pebble}. A \textit{segment} is defined as a length-$H$ sequence of observations and actions, $\{(\state_{1},\action_{1}), ...,(\state_{H}, \action_{H})\}$. Given a pair of segments $x = (A, B)$,
a user indicates which segment is preferred: $+1$ in the event that $A$ is preferred over $B$ (denoted by $A \succ B$) or $-1$ if $B$ is preferred over $A$ (denoted by $A \prec B$). 
Preference-based RL aims to train an agent to perform human-desirable behaviors according to the human user's preferences.

Our problem formulation assumes access to a \textit{crowd}, or ensemble, of users, who each assign pairwise preference labels to a sequentially-selected set $\mathcal{S}=\{x_k\}_{k=1}^{S}$, $S = |\mathcal{S}|$ of segment pairs. Each query $x_k$ takes the form $x_k = (A_k, B_k)$, where $A_k, B_k$ belong to the set $\mathcal{X}$ of available segments. In addition, we assume that a majority (e.g. at least $51\%$) of the crowd provides preference feedback according to a shared understanding of desired agent behavior. More concretely, we assume the existence of an unknown ground-truth reward $r$, and that for a majority of users, their preferences are (noisily) generated with respect to $r$. Note that this formulation allows one or more minority groups of users to potentially provide feedback according to a different reward function(s). Beyond this, we do not assume any further knowledge about the crowd (e.g. expertise, reliability). Our goal is to learn a crowd-informed preference label for each segment pair using only the preference labels given by the crowd. Formally, let ${ \{f_i\}_{i=1}^M }$ represent $M$ users of unknown reliability and error rate, where each user $i$ provides a preference label $f_i(x_k) \in \{-1, 1\}$ on each input query $x_k \in \mathcal{S}$. 
Then, let $\bm{y}$ be the vector of (unknown to the algorithm) ground-truth labels according to the true reward $r$ (i.e., in which segments with higher ground-truth reward are preferred),
$\bm{y} = [y_1, ..., y_S ]^T$. Using only the preference labels from the $M$ users on the segment pairs in $\mathcal{S}$ and without access to any ground-truth rewards or preferences, 
we seek an optimal estimate $\hat{\bm{y}} = [\hat{y}_1, ..., \hat{y}_S]^T$ of the true labels $\bm{y}$ given the crowd preference labels. We also investigate the ability to identify minority groups of users who provide feedback that consistently differs from that of the majority.

\section{Methods}


\subsection{Simulating Diverse Crowds}

To systematically evaluate the impact of user error rates
on learning from crowdsourced user preferences, we apply the Stochastic Preference Model \citep{lee2021bpref} to simulate users with mixtures of different irrationality levels:
\begin{flalign}
     P[A \succ B; \beta, \gamma] =  & \hspace{5mm} \frac{e^{\left(\beta \sum_{t=1}^{H}  \gamma^{H-t} r(\state_{t}^A, \action_{t}^A)\right)}}{e^{ \left( \beta  \sum_{t=1}^{H} \gamma^{H-t} r(\state_{t}^A, \action_{t}^A)\right)} + e^{\left( \beta \sum_{t=1}^{H} \gamma^{H-t} r(\state_{t}^B, \action_{t}^B)\right)} 
         }, \label{eq:bt_model}
\end{flalign}
\noindent where $\gamma \in (0, 1]$ is a discount factor to model myopic behavior, and $\beta$ is a rationality constant.
Note that user feedback is fully rational and deterministic as $\beta \rightarrow \infty$, whereas $\beta=0$ produces uniformly random choices. An additional parameter $\epsilon$ models users making mistakes due to accidental errors. For our analysis, each user is simulated with a random draw of parameters from the following ranges: $\gamma \in [0.98, 1.0]$, $\beta \in [0.1, 10.0]$, and $\epsilon \in [0, 0.2]$. These parameter ranges were chosen to produce diverse user error rates primarily between $15\%$ and $40\%$, which is approximately the range of human error rates observed in a prior study of human feedback in preference-based RL ($12\% - 37\%$ in \citet{brown2019extrapolating}). We then sample crowds of $M \in \{7, 11, 15\}$ users according to this model to study the effect of crowd size on crowd label learning.

\subsection{Estimating Preference Labels from a Crowd}

For each query, \alggen distills the $M$ preference labels from each crowd member to a single preference label. We discuss two methods for accomplishing this distillation. The first, majority voting, is a comparatively naive approach, while the second, the Spectral Meta Learner (SML) \citep{parisiSML}, is the method we recommend for use with \alggen. 

\newsec{Majority Vote (Baseline)}
Given the current problem formulation, in which we lack additional information about the crowd, perhaps the simplest estimate $\hat{\bm{y}}$ is the majority vote (MAJ), which labels each segment pair $x_k \in \mathcal{S}$ with the majority preference from the crowd, denoted as $\hat{\bm{y}}_{MAJ}$.

\newsec{Spectral Meta-Learner}
However, it is indeed possible to estimate a crowd label that often outperforms the majority vote in the fully unsupervised case studied here. Under the assumption that users in a crowd make independent errors, \citet{parisiSML} derived the SML:
\begin{equation}
    \hat{y}_k = \sign\left(\sum_{i=1}^M f_i(x_k) * \hat{v}_i \right),
    \label{eq:SML}
\end{equation}
\noindent where $\hat{v}_i$ are the values of $\hat{\bm{v}}$, the lead eigenvector of the empirical data covariance matrix 
of the users' binary preferences across the set of pairwise comparison queries.
As shown in \citet{parisiSML}, crowd labels derived via the SML are typically more accurate than labels from any individual crowd member or from a majority vote, as unlike other methods such as majority vote, the SML approximates a maximum likelihood estimate of the labels. In addition, \citet{parisiSML} showed that $\hat{\bm{v}}$ can rank the users by their estimated error rates, which approximate the ground truth error rates. We denote the labels estimated by SML as $\hat{\bm{y}}_{SML}$. \citet{parisiSML} also analyze the performance of SML in the presence of a \textit{cartel}, a small group of users who attempt to steer the overall ensemble solution away from the (unknown) ground truth, and proved that (1) SML is robust to the presence of cartels, and that (2) the weights of the eigenvector $\hat{\bm{v}}$ can be used to identify the presence of cartels in the crowd. Here, we redefine the notion of a cartel as a minority group of users who provide feedback that consistently differs from that of the majority, for example users who provide feedback towards goals that are different from the goals of the majority. Notably, to our knowledge, the SML has not previously been applied toward preference-based learning or RL.

\subsection{Crowdsourced Preference Learning for RL}
\alggen (see Algorithm 1 in the Supplemental Material) leverages the crowd-aggregated preference labels $\hat{\bm{y}} = \hat{\bm{y}}_{SML}$ for all instances in $\mathcal{S}$. We denote via $\mathcal{D}$ the dataset of all pairs $\{(x, y)\}$, where $x \in \mathcal{S}$ and $y$ is the corresponding crowd-aggregated label in $\hat{\bm{y}}$. To learn useful behaviors given these aggregated labels, we assume that the majority of users (though not necessarily all users) have a shared understanding of desired behavior, which can be quantified in terms of an underlying reward function. This is done by iterating between the following two steps: (1) a reward function learning step, where we optimize the learned reward via supervised learning on the preference labels $\hat{\bm{y}}$ given over behaviors generated by the policy $\pi$, and (2) a policy learning step, where the agent performs environment rollouts and optimizes its policy against the current reward function estimate. Segment pairs for preference queries are selected according to the disagreement among an ensemble of reward predictors, similarly to \citet{lee2021bpref}. We then learn a reward model $\widehat{r}_\psi$, modeled as a neural network with parameters $\psi$, that predicts the user preferences as follows:
\begin{align}
  P_\psi[A\succ\ B] = \frac{\exp\sum_{t} \rewmodel(\state_{t}^A, \action_{t}^A)}{\sum_{i\in \{A, B\}} \exp\sum_{t} \rewmodel(\state_{t}^i, \action_{t}^i)}\label{eq:pref_model}.
\end{align}
The reward function  $\widehat{r}_\psi$ is then updated by minimizing the following cross-entropy loss:
\begin{align}
  \mathcal{L}^{\tt Reward} =  -\expec_{(A,B,y)\sim \mathcal{D}} \Big[ & \mathbbm{1}_{[y = -1]}\log P_\psi[B\succ A] + \mathbbm{1}_{[y = 1]} \log P_\psi[A\succ B]\Big]\label{eq:reward-bce},
\end{align}
where $\mathbbm{1}_{[\cdot]}$ denotes an indicator variable.
Given the learned reward function $\widehat{r}_\psi$, a policy $\pi$ can be learned using any RL algorithm. In this work, we use PrefPPO \citep{lee2021bpref}, instantiating \alggen as \alg. PrefPPO uses Proximal Policy Optimization (PPO), a state-of-the-art on-policy RL algorithm \citep{schulman2017proximal}. As recommended in \citet{lee2021bpref}, we use on-policy methods to reduce the effects of non-stationarity induced by learning a reward function and policy simultaneously.

\section{Experiments}

\begin{figure*}[t!]
\centering
\includegraphics[scale=0.3]{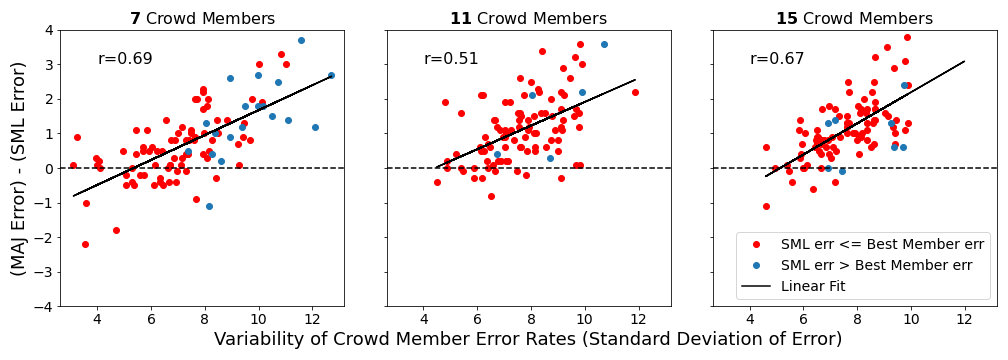}
\vskip -0.1 true in
\caption{Difference in MAJ and SML preference prediction error rates for different levels of variability (standard deviation) in user error rates. 
Values are calculated across 100 randomly-sampled crowds of 7, 11 and 15 simulated users for the \texttt{Walker-walk} environment. The horizontal dashed line at $y=0$ indicates where the MAJ and SML error rates are the same; points above $y=0$ indicate that the SML labels have lower error than the MAJ labels. Red dots indicate where SML outperforms the best crowd member; blue indicates where the best crowd member outperforms SML.}
\label{fig:ensemble_error}
\end{figure*}

\subsection{Experiment Setup}
We conduct experiments to evaluate the proposed approach to crowdsourcing human feedback for preference learning. Our experiments conduct RL from human preferences using the PrefPPO implementation in \citet{lee2021bpref}.
These experiments use the \texttt{Walker-walk}, \texttt{Quadruped-walk}, and \texttt{Cheetah-run} environments from the DMControl suite \citep{dmcontrol}. For each environment, we report results for \alg-SML (SML preference estimation) and the following comparisons: \alg-MAJ (in which the majority vote is used instead of the SML estimate) and \alg-Oracle (in which the preference is given to the segment with the highest ground-truth environment reward); in addition, we compare \alg's estimated preference labels to those from the best member of each crowd.  We perform 10 total runs of each algorithm comparison in each environment for 4M environment steps. Methods are compared via the mean and standard error over 6 runs, where the top 2 and bottom 2 runs are omitted (measured by episode return at 4M environment steps) to reduce the impact of outliers (similar to the InterQuartile Mean, IQM \citep{Agarwal2021}). All remaining hyperparameters are the same as reported in \citet{lee2021bpref} and listed in the Supplemental Materials.

\subsection{Results}

We first study the effect of crowd size on the performance of crowdsourced preference learning by simulating 100 different random crowd configurations at each of three different crowd sizes ($M \in \{7, 11, 15\}$) according to the Stochastic Preference Model in Equation~\eqref{eq:bt_model} for the \texttt{Walker-Walk} environment.  Figure~\ref{fig:ensemble_error} depicts the 
difference in errors between the MAJ and SML preference estimates versus the diversity in users' error rates with respect to the ground-truth reward (quantified via the standard deviation of user error across the crowd) for each simulated crowd size. To calculate these errors for each simulated crowd, we ran \alg using Oracle preference labels to generate sets of segment pairs for preference labeling and reward learning. Then, we use these segment pairs and their ground-truth labels to construct the error rates for all individual users in the crowd and the error rates corresponding to the MAJ and SML aggregated labels. 
Since results were similar for the other environments tested, we show results for only the \texttt{Walker-walk} environment due to space constraints. We see a positive correlation between the error rates of users in the crowd and the performance of SML compared to MAJ. In particular, as the user variability increases, the better the SML labels perform compared to the MAJ labels, indicating that SML is filtering out inconsistent preference feedback from users with higher error rates, and SML is aggregating crowd decisions more consistently than MAJ. We also see that SML outperforms MAJ in nearly all cases ($\approx 90\%$ of points above $y=0$), suggesting that SML captures the crowd preferences more accurately than MAJ. In addition, the SML labels have lower errors than the best crowd member's labels in nearly all cases (red dots). Finally, across all crowd sizes, the spread of the difference in MAJ errors and SML errors is fairly consistent (most points fall between $[0, 4]\%$), suggesting that SML labels can outperform MAJ labels with most crowd configurations. While this improvement may seem small, it occurs during every reward learning iteration; we believe that this leads to compounding benefits during reward learning, and consequently, during agent policy learning. 

\begin{figure*}[t!]
\centering
    \includegraphics[scale=0.26]{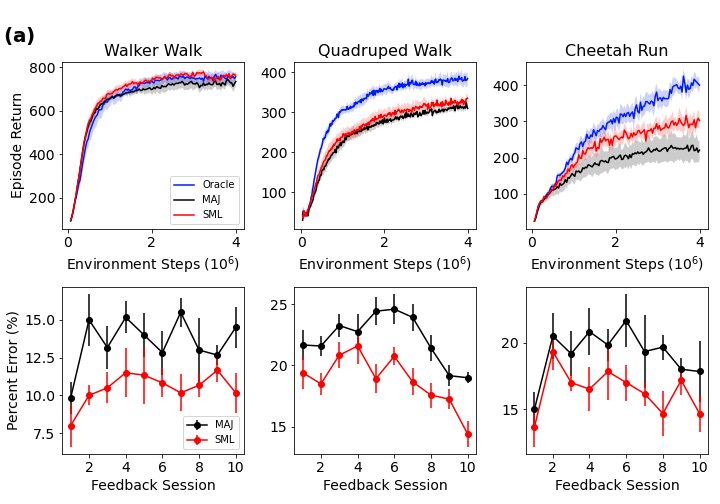}
    \includegraphics[scale=0.26]{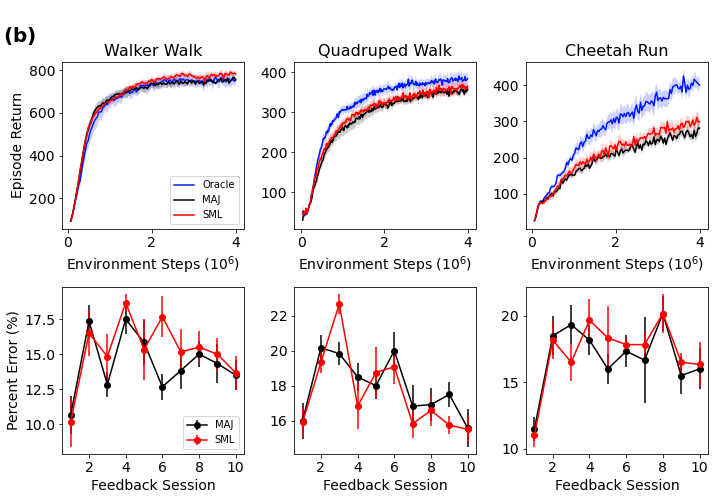}
\vskip -0.1 true in
\caption{(Top row) Comparison of \alg training curves with SML, MAJ and Oracle labeling across two different crowd configurations: [(a), left] a crowd for which SML is expected to outperform MAJ, and [(b), right] a crowd for which SML is expected to perform similarly to MAJ across the three different environments (\texttt{Walker-walk}, \texttt{Quadruped-walk} and \texttt{Cheetah-run}). (Bottom row) Comparison of MAJ and SML label prediction errors at each feedback iteration. Each plot shows the mean $\pm$ standard error of 6 out of 10 runs (the top and bottom 2 runs are omitted to reduce the effect of outliers, as detailed in the Experiment Setup).} 
\label{fig:rewards}
\end{figure*}

We next conduct experiments with \alg 
to determine if the improved label error from SML observed in Figure~\ref{fig:ensemble_error} impacts agent performance in a meaningful way. For this analysis, we select a fixed crowd size of $M=15$ users and sample two different crowd configurations: (1) Figure \ref{fig:rewards}a studies a crowd for which SML outperforms MAJ by $\approx 4\%$ in prediction error (following the analysis in Figure~\ref{fig:ensemble_error}), and (2) Figure \ref{fig:rewards}b studies a crowd for which SML performs similarly to MAJ in prediction error (similarly from the analysis in Figure~\ref{fig:ensemble_error}). In addition, in the bottom row of Figure~\ref{fig:rewards}, we plot the error rates of the MAJ and SML label predictions relative to the ground-truth labels during each crowd feedback iteration for all three tested environments.
We see that \alg with SML label aggregation (\alg-SML) outperforms \alg with MAJ labels (\alg-MAJ) in terms of trajectory return for all tested environments, with \alg achieving close to Oracle performance for \texttt{Walker-walk} and \texttt{Quadruped-walk}. In addition, the SML label errors are lower than the MAJ label errors at each feedback iteration in most cases for each environment. The lower SML label errors imply that there is a compounding benefit to using SML labels rather than MAJ labels for reward learning; in early feedback iterations, the learned reward has lower error and thus is closer to the Oracle reward function compared to MAJ, improving agent training. We see a similar, but weaker, trend in Figure \ref{fig:rewards}b, where SML is expected to yield a similar label accuracy to MAJ for the given crowd configuration. There is generally less spread in episode returns between SML and MAJ for the three tested environments (Figure \ref{fig:rewards}b, top), and we see that the crowd feedback errors are closer between the methods during all crowd feedback iterations (Figure \ref{fig:rewards}b, bottom), indicating only minimal benefits of SML labels over MAJ labels.

\begin{figure}[t!]
\centering
\includegraphics[scale=0.42]{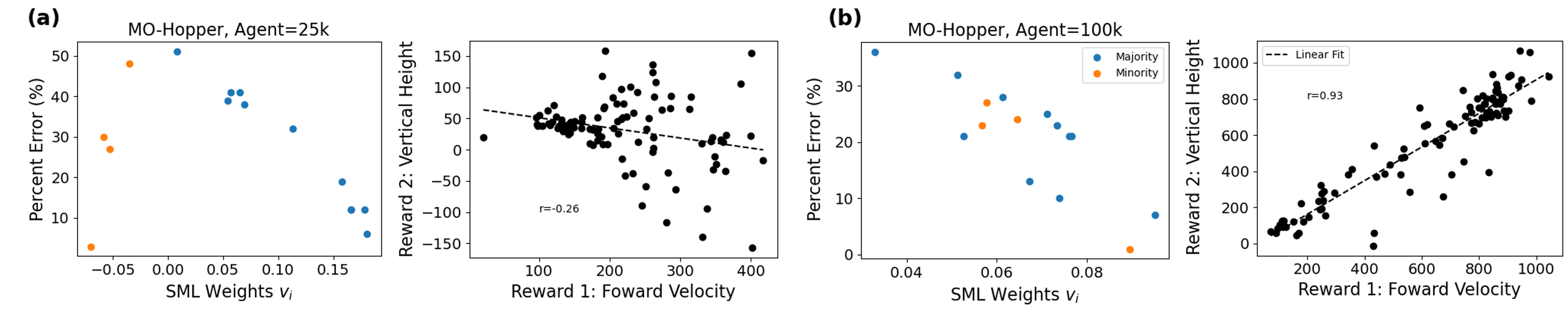}
\vskip -0.1 true in
\caption{SML weights and agent performance in \texttt{MO-Hopper}. (a) 25K training steps.  Left: Scatter plot of crowd SML weights $\hat{v}_i$ with corresponding user error rates. Right: scatter plot of returns of 100 rollout trajectories with two reward functions, (1) Forward and (2) Vertical Height.
(b) Similar to (a) with 100K training steps. At 25K training steps (a), we observe a reliable separation between the minority and majority groups and weakly negatively correlated returns in the two reward functions, indicating that the agent has not yet learned to optimize both objectives. In contrast, at 100K training steps (b), the agent has learned to satisfy both sets of users, as indicated by the large positive correlation in returns, and thus the SML weights no longer separate into two groups. 
}
\label{fig:ranking_majority_minority}
\end{figure}

Finally, we conduct an experiment to test the ability of SML to detect the presence of minority group feedback based purely on the crowd preference labels. For this experiment, we use the \texttt{MO-Hopper} environment found in Multi-Objective Gymnasium \citep{felten_toolkit_2023}. This environment encodes two reward functions according to two different objectives: (1) forward velocity, which encourages the agent to run forward and (2) vertical height, which encourages the agent to jump high. We simulate a crowd of 15 users, 11 of whom assign preference feedback according to Objective 1 (forward velocity) while the remainder assign feedback according to Objective 2 (vertical height). We aim to determine if the SML weights can identify the presence of minority groups within the crowd. To obtain the initial set of trajectories for crowd labeling, we first train an agent via PPO using \textit{stable-baselines3} \citep{raffin2021} on a linearized reward that averages the two reward functions, and evaluate the learned model after both 25K and 100K training frames. At both of these checkpoints, we collect 100 agent rollout trajectories, sample 100 trajectory pairs for crowd labeling, and calculate the SML labels and each user's SML weight $\hat{v}_i$. Results are shown in Figure \ref{fig:ranking_majority_minority}. First, in Figure \ref{fig:ranking_majority_minority}a (left), we see that at 25K training steps, there is a separation of the minority (orange) and majority (blue) crowd SML weights, 
with an approximate threshold at $\hat{v}_i=0$ separating the two groups. We also see (Figure \ref{fig:ranking_majority_minority}a, right) that the trajectory returns 
are weakly negatively correlated across the two objectives, indicating that the agent has not yet learned to optimize both objectives simultaneously (e.g. preferring to run forward instead of jump for some trajectories, and vice-versa). At 100K training steps (Figure \ref{fig:ranking_majority_minority}b), we see a different result: the agent now can optimize behaviors according to both objectives (as indicated by high positive correlation in the two returns), while the SML weights can no longer separate the presence of minority and majority groups. This is because the agent behaviors now satisfy the objectives of both the minority and majority groups, so that even though the two groups assign feedback according to different reward functions, there is no longer any disagreement in crowd preferences.

\begin{figure}[t!]
\centering
\includegraphics[scale=0.475]{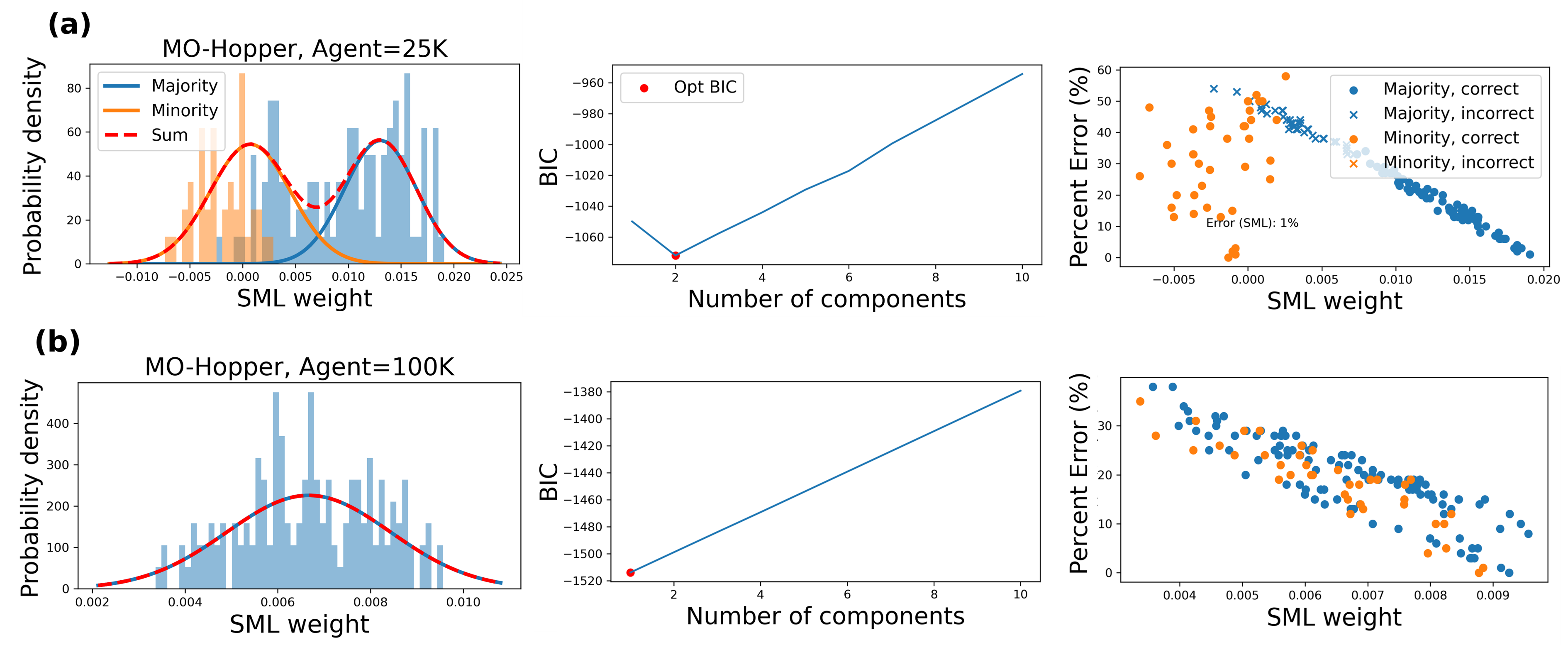}
\vskip -0.1 true in
\caption{Unsupervised clustering of users based purely on crowd feedback. We consider a crowd of 150 workers with a 110/40 majority/minority split in the \texttt{MO-Hopper} environment. (a) For 25K agent training steps, we see (left) a histogram of SML weights $\hat{v}_i$ overlaid with a fitted GMM, (middle) the BIC values used to automatically infer the number of clusters, and (right) a scatter plot of SML weights vs. ground-truth error showing the accuracy of the inferred grouping. Colors indicate the ground truth grouping (blue for majority, orange for minority). 
(b) Similar to (a) with 100K training steps. 
We see that for the agent at 25K training steps (a), a GMM with two Gaussian components reliably models the two groups using only the SML weights $\hat{v}_i$, automatically inferring the number of clusters and performing unsupervised cluster assignment. At 100K training frames (b), the agent can satisfy both sets of users, and so the cluster analysis no longer infers any minority.}
\vskip -0.1 true in
\label{fig:gmm_cluster_assignment}
\end{figure}

Figure \ref{fig:gmm_cluster_assignment} further explores this phenomenon when working with very large crowds (150 users, 110/40 split into majority/minority groups) for the \texttt{MO-Hopper} environment. We fit a one-dimensional Gaussian Mixture Model (GMM) to the SML weights $\hat{\bm{v}}$ to automatically cluster users into majority and minority groups in the absence of ground-truth information (Figure \ref{fig:gmm_cluster_assignment}, left column), using the Bayesian Information Criterion \citep{schwarz1978estimating} (BIC; Figure \ref{fig:gmm_cluster_assignment}, middle column) to infer the optimal number of Gaussian distributions in both cases (25K and 100K training steps). We see that at 25K training steps, there is a fairly reliable separation between the minority and majority groups based purely on the SML weights $\hat{\bm{v}}$, fitting two Gaussian distributions (one per group).
Meanwhile, for the agent at 100K training steps, we do not see any meaningful user separation, as the BIC indicates only one user cluster. We then conducted a proof-of-concept analysis (Figure \ref{fig:gmm_cluster_assignment}, right column) using the user groupings \textit{inferred} by the GMM
to separately estimate the crowd preference labels for the inferred minority and majority classes via SML label aggregation. We found that the aggregated crowd labels have very low error compared to ground-truth feedback ($\approx 1\%$ for Minority, 0$\%$ for Majority), suggesting that a reward learned with these labels
would be practically identical to one learned with oracle labels. 
These results suggest that \alggen with SML crowd label inference could be used to infer the number of user groups, cluster users via their objectives, and learn reward functions and corresponding behaviors specific to each user cluster.

\section{Discussion}
This study proposes the \alggen framework for training AI agents to integrate feedback from many users. Our evaluation of \alg indicates that we can use methods derived from unsupervised ensemble learning (SML, \citet{parisiSML}) to effectively aggregate user preference feedback across diverse crowds to learn RL policies without any ground-truth user information. 
 Furthermore, our evaluation suggests that \alg can learn a reward from crowd-aggregated labels that is more accurate than that derived from any single user's feedback. Finally, we show the possibility of detecting minority feedback clusters in an unsupervised manner, potentially enabling a deeper understanding of model behavior, e.g. via detecting multiple user viewpoints, characterizing the associated objectives, and mitigating the effects of malicious user data. 

We showed that SML labels outperform MAJ labels in nearly all cases across different crowd sizes (Figure \ref{fig:ensemble_error}) and different tasks (Figure \ref{fig:rewards}). 
Calculating SML weights is computationally very fast, only requiring a few lines of code in modern programming languages (e.g. Python). This indicates that simple code modifications can result in significant performance gains when working with crowdsourced preferences. 

We further evaluate \alg on a multi-objective robotics task, showing that \alg can identify minority groups in the crowd and demonstrating the method's potential to learn minority reward functions (Figure~\ref{fig:gmm_cluster_assignment}). How to best incorporate this knowledge into the agent training loop remains an open question; our framework could lend itself to multiple approaches, e.g. learning a utility function that attempts to balance the multiple users' desired behavior, learning personalized reward functions for specific user groups, or allowing human users to assign weights to each reward function. In future work, we hope to explore how to robustly account for multiple learned objectives across these various scenarios. 

Several limitations of our approach are important to discuss. First, each user in the crowd must provide a preference label for each preference query. Future work will focus on the possibility of using \alggen when user feedback is sparse. In addition, this study uses only simulated human feedback, which allows for a systematic evaluation of \alggen and an analysis of its properties in a controlled setting. Future work will evaluate this method in a human subjects study.

\bibliography{main}

\begin{thebibliography}{31}
\providecommand{\natexlab}[1]{#1}
\providecommand{\url}[1]{\texttt{#1}}
\expandafter\ifx\csname urlstyle\endcsname\relax
  \providecommand{\doi}[1]{DOI: #1}\else
  \providecommand{\doi}{DOI: \begingroup \urlstyle{rm}\Url}\fi

\bibitem[Agarwal et~al.(2021)Agarwal, Schwarzer, Castro, Courville, and Bellemare]{Agarwal2021}
Rishabh Agarwal, Max Schwarzer, Pablo~Samuel Castro, Aaron~C Courville, and Marc Bellemare.
\newblock Deep reinforcement learning at the edge of the statistical precipice.
\newblock In M.~Ranzato, A.~Beygelzimer, Y.~Dauphin, P.S. Liang, and J.~Wortman Vaughan (eds.), \emph{Advances in Neural Information Processing Systems}, volume~34, pp.\  29304--29320. Curran Associates, Inc., 2021.
\newblock URL \url{https://proceedings.neurips.cc/paper_files/paper/2021/file/f514cec81cb148559cf475e7426eed5e-Paper.pdf}.

\bibitem[Bai et~al.(2022)Bai, Jones, Ndousse, Askell, Chen, DasSarma, Drain, Fort, Ganguli, Henighan, Joseph, Kadavath, Kernion, Conerly, El-Showk, Elhage, Hatfield-Dodds, Hernandez, Hume, Johnston, Kravec, Lovitt, Nanda, Olsson, Amodei, Brown, Clark, McCandlish, Olah, Mann, and Kaplan]{bai2022traininghelpfulharmlessassistant}
Yuntao Bai, Andy Jones, Kamal Ndousse, Amanda Askell, Anna Chen, Nova DasSarma, Dawn Drain, Stanislav Fort, Deep Ganguli, Tom Henighan, Nicholas Joseph, Saurav Kadavath, Jackson Kernion, Tom Conerly, Sheer El-Showk, Nelson Elhage, Zac Hatfield-Dodds, Danny Hernandez, Tristan Hume, Scott Johnston, Shauna Kravec, Liane Lovitt, Neel Nanda, Catherine Olsson, Dario Amodei, Tom Brown, Jack Clark, Sam McCandlish, Chris Olah, Ben Mann, and Jared Kaplan.
\newblock Training a helpful and harmless assistant with reinforcement learning from human feedback, 2022.
\newblock URL \url{https://arxiv.org/abs/2204.05862}.

\bibitem[Brown et~al.(2019)Brown, Goo, Nagarajan, and Niekum]{brown2019extrapolating}
Daniel Brown, Wonjoon Goo, Prabhat Nagarajan, and Scott Niekum.
\newblock Extrapolating beyond suboptimal demonstrations via inverse reinforcement learning from observations.
\newblock In \emph{Proceedings of the International Conference on Machine Learning}, pp.\  783--792, 2019.

\bibitem[Casper et~al.(2023)Casper, Davies, Shi, Gilbert, Scheurer, Rando, Freedman, Korbak, Lindner, Freire, et~al.]{casper2023open}
Stephen Casper, Xander Davies, Claudia Shi, Thomas~Krendl Gilbert, J{\'e}r{\'e}my Scheurer, Javier Rando, Rachel Freedman, Tomasz Korbak, David Lindner, Pedro Freire, et~al.
\newblock Open problems and fundamental limitations of reinforcement learning from human feedback.
\newblock \emph{Transactions on Machine Learning Research}, 2023.

\bibitem[Chakraborty et~al.(2024)Chakraborty, Qiu, Yuan, Koppel, Manocha, Huang, Bedi, and Wang]{chakraborty2024maxminrlhf}
Souradip Chakraborty, Jiahao Qiu, Hui Yuan, Alec Koppel, Dinesh Manocha, Furong Huang, Amrit Bedi, and Mengdi Wang.
\newblock Maxmin-{RLHF}: Alignment with diverse human preferences.
\newblock In \emph{Forty-first International Conference on Machine Learning}, 2024.
\newblock URL \url{https://openreview.net/forum?id=8tzjEMF0Vq}.

\bibitem[Chen et~al.(2013)Chen, Bennett, Collins-Thompson, and Horvitz]{Chen2013}
Xi~Chen, Paul~N. Bennett, Kevyn Collins-Thompson, and Eric Horvitz.
\newblock Pairwise ranking aggregation in a crowdsourced setting.
\newblock In \emph{Proceedings of the Sixth ACM International Conference on Web Search and Data Mining}, WSDM '13, pp.\  193–202, New York, NY, USA, 2013. Association for Computing Machinery.
\newblock ISBN 9781450318693.
\newblock \doi{10.1145/2433396.2433420}.
\newblock URL \url{https://doi.org/10.1145/2433396.2433420}.

\bibitem[Christiano et~al.(2017)Christiano, Leike, Brown, Martic, Legg, and Amodei]{christiano2017deep}
Paul~F Christiano, Jan Leike, Tom Brown, Miljan Martic, Shane Legg, and Dario Amodei.
\newblock Deep reinforcement learning from human preferences.
\newblock \emph{Advances in neural information processing systems}, 30, 2017.

\bibitem[Durmus et~al.(2024)Durmus, Nguyen, Liao, Schiefer, Askell, Bakhtin, Chen, Hatfield-Dodds, Hernandez, Joseph, Lovitt, McCandlish, Sikder, Tamkin, Thamkul, Kaplan, Clark, and Ganguli]{durmus2024measuringrepresentationsubjectiveglobal}
Esin Durmus, Karina Nguyen, Thomas~I. Liao, Nicholas Schiefer, Amanda Askell, Anton Bakhtin, Carol Chen, Zac Hatfield-Dodds, Danny Hernandez, Nicholas Joseph, Liane Lovitt, Sam McCandlish, Orowa Sikder, Alex Tamkin, Janel Thamkul, Jared Kaplan, Jack Clark, and Deep Ganguli.
\newblock Towards measuring the representation of subjective global opinions in language models, 2024.
\newblock URL \url{https://arxiv.org/abs/2306.16388}.

\bibitem[Felten et~al.(2023)Felten, Alegre, Now{\'e}, Bazzan, Talbi, Danoy, and Silva]{felten_toolkit_2023}
Florian Felten, Lucas~N. Alegre, Ann Now{\'e}, Ana L.~C. Bazzan, El~Ghazali Talbi, Gr{\'e}goire Danoy, and Bruno C.~{\relax da} Silva.
\newblock A toolkit for reliable benchmarking and research in multi-objective reinforcement learning.
\newblock In \emph{Proceedings of the 37th Conference on Neural Information Processing Systems ({NeurIPS} 2023)}, 2023.

\bibitem[Lee et~al.(2021{\natexlab{a}})Lee, Smith, Dragan, and Abbeel]{lee2021bpref}
Kimin Lee, Laura Smith, Anca Dragan, and Pieter Abbeel.
\newblock B-pref: Benchmarking preference-based reinforcement learning.
\newblock In \emph{Thirty-fifth Conference on Neural Information Processing Systems Datasets and Benchmarks Track (Round 1)}, 2021{\natexlab{a}}.
\newblock URL \url{https://openreview.net/forum?id=ps95-mkHF_}.

\bibitem[Lee et~al.(2021{\natexlab{b}})Lee, Smith, and Abbeel]{lee2021pebble}
Kimin Lee, Laura~M Smith, and Pieter Abbeel.
\newblock {PEBBLE}: Feedback-efficient interactive reinforcement learning via relabeling experience and unsupervised pre-training.
\newblock In \emph{International Conference on Machine Learning}, pp.\  6152--6163, 2021{\natexlab{b}}.

\bibitem[Li et~al.(2016)Li, Wang, Zheng, and Franklin]{Li2016}
Guoliang Li, Jiannan Wang, Yudian Zheng, and Michael~J. Franklin.
\newblock Crowdsourced data management: A survey.
\newblock \emph{IEEE Transactions on Knowledge and Data Engineering}, 28\penalty0 (9):\penalty0 2296--2319, 2016.
\newblock \doi{10.1109/TKDE.2016.2535242}.

\bibitem[Nakano et~al.(2022)Nakano, Hilton, Balaji, Wu, Ouyang, Kim, Hesse, Jain, Kosaraju, Saunders, Jiang, Cobbe, Eloundou, Krueger, Button, Knight, Chess, and Schulman]{nakano2022webgptbrowserassistedquestionansweringhuman}
Reiichiro Nakano, Jacob Hilton, Suchir Balaji, Jeff Wu, Long Ouyang, Christina Kim, Christopher Hesse, Shantanu Jain, Vineet Kosaraju, William Saunders, Xu~Jiang, Karl Cobbe, Tyna Eloundou, Gretchen Krueger, Kevin Button, Matthew Knight, Benjamin Chess, and John Schulman.
\newblock Webgpt: Browser-assisted question-answering with human feedback, 2022.
\newblock URL \url{https://arxiv.org/abs/2112.09332}.

\bibitem[Ouyang et~al.(2022)Ouyang, Wu, Jiang, Almeida, Wainwright, Mishkin, Zhang, Agarwal, Slama, Ray, Schulman, Hilton, Kelton, Miller, Simens, Askell, Welinder, Christiano, Leike, and Lowe]{Ouyang2022}
Long Ouyang, Jeffrey Wu, Xu~Jiang, Diogo Almeida, Carroll Wainwright, Pamela Mishkin, Chong Zhang, Sandhini Agarwal, Katarina Slama, Alex Ray, John Schulman, Jacob Hilton, Fraser Kelton, Luke Miller, Maddie Simens, Amanda Askell, Peter Welinder, Paul~F Christiano, Jan Leike, and Ryan Lowe.
\newblock Training language models to follow instructions with human feedback.
\newblock In S.~Koyejo, S.~Mohamed, A.~Agarwal, D.~Belgrave, K.~Cho, and A.~Oh (eds.), \emph{Advances in Neural Information Processing Systems}, volume~35, pp.\  27730--27744. Curran Associates, Inc., 2022.
\newblock URL \url{https://proceedings.neurips.cc/paper_files/paper/2022/file/b1efde53be364a73914f58805a001731-Paper-Conference.pdf}.

\bibitem[Parisi et~al.(2014)Parisi, Strino, Nadler, and Kluger]{parisiSML}
Fabio Parisi, Francesco Strino, Boaz Nadler, and Yuval Kluger.
\newblock Ranking and combining multiple predictors without labeled data.
\newblock \emph{Proceedings of the National Academy of Sciences}, 111\penalty0 (4):\penalty0 1253--1258, 2014.
\newblock \doi{10.1073/pnas.1219097111}.
\newblock URL \url{https://www.pnas.org/doi/abs/10.1073/pnas.1219097111}.

\bibitem[Poddar et~al.(2024)Poddar, Wan, Ivison, Gupta, and Jaques]{poddar2024personalizing}
Sriyash Poddar, Yanming Wan, Hamish Ivison, Abhishek Gupta, and Natasha Jaques.
\newblock Personalizing reinforcement learning from human feedback with variational preference learning.
\newblock \emph{arXiv preprint arXiv:2408.10075}, 2024.

\bibitem[Prelec et~al.(2017)Prelec, Seung, and McCoy]{Prelec2017Nature}
Dražen Prelec, H.~Sebastian Seung, and John McCoy.
\newblock A solution to the single-question crowd wisdom problem.
\newblock \emph{Nature}, 541\penalty0 (7638):\penalty0 532--535, 2017.
\newblock URL \url{https://EconPapers.repec.org/RePEc:nat:nature:v:541:y:2017:i:7638:d:10.1038_nature21054}.

\bibitem[Rafailov et~al.(2023)Rafailov, Sharma, Mitchell, Manning, Ermon, and Finn]{rafailov2023direct}
Rafael Rafailov, Archit Sharma, Eric Mitchell, Christopher~D Manning, Stefano Ermon, and Chelsea Finn.
\newblock Direct preference optimization: Your language model is secretly a reward model.
\newblock \emph{Advances in Neural Information Processing Systems}, 36:\penalty0 53728--53741, 2023.

\bibitem[Raffin et~al.(2021)Raffin, Hill, Gleave, Kanervisto, Ernestus, and Dormann]{raffin2021}
Antonin Raffin, Ashley Hill, Adam Gleave, Anssi Kanervisto, Maximilian Ernestus, and Noah Dormann.
\newblock Stable-baselines3: Reliable reinforcement learning implementations.
\newblock \emph{Journal of Machine Learning Research}, 22\penalty0 (268):\penalty0 1--8, 2021.
\newblock URL \url{http://jmlr.org/papers/v22/20-1364.html}.

\bibitem[Ramesh et~al.(2024)Ramesh, Hu, Chaimalas, Mehta, and Sessa]{ramesh2024group}
Shyam~Sundhar Ramesh, Yifan Hu, Iason Chaimalas, Viraj Mehta, and Pier~Giuseppe Sessa.
\newblock Group robust preference optimization in reward-free rlhf.
\newblock In \emph{38th Annual Conference on Neural Information Processing Systems (NeurIPS 2024)}, 2024.

\bibitem[Schulman et~al.(2017)Schulman, Wolski, Dhariwal, Radford, and Klimov]{schulman2017proximal}
John Schulman, Filip Wolski, Prafulla Dhariwal, Alec Radford, and Oleg Klimov.
\newblock Proximal policy optimization algorithms.
\newblock \emph{arXiv preprint arXiv:1707.06347}, 2017.

\bibitem[Schwarz(1978)]{schwarz1978estimating}
Gideon Schwarz.
\newblock Estimating the dimension of a model.
\newblock \emph{The Annals of Statistics}, pp.\  461--464, 1978.

\bibitem[Siththaranjan et~al.(2023)Siththaranjan, Laidlaw, and Hadfield-Menell]{siththaranjan2023distributional}
Anand Siththaranjan, Cassidy Laidlaw, and Dylan Hadfield-Menell.
\newblock Distributional preference learning: Understanding and accounting for hidden context in {RLHF}.
\newblock \emph{arXiv preprint arXiv:2312.08358}, 2023.

\bibitem[Surowiecki(2005)]{Surowiecki2005}
James Surowiecki.
\newblock \emph{The Wisdom of Crowds}.
\newblock Anchor, 2005.
\newblock ISBN 0385721706.

\bibitem[Sutton \& Barto(2018)Sutton and Barto]{sutton2018reinforcement}
Richard~S Sutton and Andrew~G Barto.
\newblock \emph{Reinforcement learning: An introduction}.
\newblock MIT press, 2018.

\bibitem[Tassa et~al.(2018)Tassa, Doron, Muldal, Erez, Li, de~Las~Casas, Budden, Abdolmaleki, Merel, Lefrancq, Lillicrap, and Riedmiller]{dmcontrol}
Yuval Tassa, Yotam Doron, Alistair Muldal, Tom Erez, Yazhe Li, Diego de~Las~Casas, David Budden, Abbas Abdolmaleki, Josh Merel, Andrew Lefrancq, Timothy~P. Lillicrap, and Martin~A. Riedmiller.
\newblock Deepmind control suite.
\newblock \emph{Computing Research Repository (CoRR)}, abs/1801.00690, 2018.
\newblock URL \url{http://arxiv.org/abs/1801.00690}.

\bibitem[Torne et~al.(2023)Torne, Balsells, Wang, Desai, Chen, Agrawal, and Gupta]{torne2023breadcrumbs}
Marcel Torne, Max Balsells, Zihan Wang, Samedh Desai, Tao Chen, Pulkit Agrawal, and Abhishek Gupta.
\newblock Breadcrumbs to the goal: Goal-conditioned exploration from human-in-the-loop feedback.
\newblock \emph{arXiv preprint arXiv:2307.11049}, 2023.

\bibitem[Wilde et~al.(2021)Wilde, Biyik, Sadigh, and Smith]{wilde2021learning}
Nils Wilde, Erdem Biyik, Dorsa Sadigh, and Stephen~L. Smith.
\newblock Learning reward functions from scale feedback.
\newblock In \emph{5th Annual Conference on Robot Learning}, 2021.
\newblock URL \url{https://openreview.net/forum?id=udFuJTvlhsJ}.

\bibitem[Yao et~al.(2025)Yao, Cai, Chuang, Yang, Jiang, Yang, and Hu]{yao2025no}
Binwei Yao, Zefan Cai, Yun-Shiuan Chuang, Shanglin Yang, Ming Jiang, Diyi Yang, and Junjie Hu.
\newblock No preference left behind: Group distributional preference optimization.
\newblock In \emph{The Thirteenth International Conference on Learning Representations}, 2025.

\bibitem[Zhang \& Kashima(2023)Zhang and Kashima]{Zhang2023}
Guoxi Zhang and Hisashi Kashima.
\newblock Batch reinforcement learning from crowds.
\newblock In Massih-Reza Amini, St{\'e}phane Canu, Asja Fischer, Tias Guns, Petra Kralj~Novak, and Grigorios Tsoumakas (eds.), \emph{Machine Learning and Knowledge Discovery in Databases}, pp.\  38--51, Cham, 2023. Springer Nature Switzerland.
\newblock ISBN 978-3-031-26412-2.

\bibitem[Zhao et~al.(2024)Zhao, Dang, and Grover]{zhao2024group}
Siyan Zhao, John Dang, and Aditya Grover.
\newblock Group preference optimization: Few-shot alignment of large language models.
\newblock In \emph{The Twelfth International Conference on Learning Representations}, 2024.

\end{thebibliography}
\bibliographystyle{rlj}

\beginSupplementaryMaterials

\section*{Full \alggen Algorithm}

The full \alggen algorithm is specified in Algorithm~\ref{alg:overview}, where red text indicates differences from the framework in ~\citet{lee2021bpref}.

\begin{algorithm}[h!]
\caption{\alggen: Preference-based RL with reward learning \textcolor{red}{from crowds}} \label{alg:overview}
\begin{algorithmic}[1]
\Require frequency of crowd feedback $K$
\Require number of queries $N_{\tt query}$ per feedback session
\Require \textcolor{red}{crowd size $M$}
\Require \textcolor{red}{crowdsource method, one of (``\texttt{MAJ}'', ``\texttt{SML}'')}
\State Initialize parameters of policy $\policy_\phi$, reward model $\rewmodel$, preference dataset $\mathcal{D} \leftarrow \emptyset$, and buffer $\mathcal{B} \leftarrow \emptyset$
\State {{\textsc{// Exploration phase}}}
\State $\mathcal{B}, \policy_\phi \leftarrow\texttt{EXPLORE}()$ according to policy $\pi_\phi$
\For{each iteration}
\State {{\textsc{// Reward learning}}}
\If{iteration \% $K == 0$} 
\For{$m$ in $1\ldots N_{\tt query}$}
\State Sample pair of segments $(A, B)$
\State $\hat{\bm{y}}_{1:M}$ $\leftarrow$ query \textcolor{red}{all $M$ users in crowd}
\EndFor
\If{\textcolor{red}{crowd method == ``\texttt{SML}''}}
\State \textcolor{red}{Compute \texttt{SML} ranking vector $\hat{\bm{v}}_i$}
\State \textcolor{red}{Compute $\hat{\bm{y}}_{SML}$ according to (\ref{eq:SML})}
\State \textcolor{red}{$\hat{\bm{y}} \leftarrow \hat{\bm{y}}_{SML}$}
\Else{ \textcolor{red}{Compute \texttt{MAJ} label estimate $\hat{\bm{y}}_{MAJ}$}}
\State \textcolor{red}{$\hat{\bm{y}} \leftarrow \hat{\bm{y}}_{MAJ}$}
\EndIf
\State $\mathcal{D} \leftarrow \mathcal{D}\cup \{(A,B,\textcolor{red}{\hat{\bm{y}}})\}$
\For{each gradient step}
\State Sample batch $\{(A, B,\textcolor{red}
{\hat{\bm{y}}})_j\}_{j=1}^{|\mathcal{D}|}\sim\mathcal{D}$ 
\State Optimize $\mathcal{L}^{\tt Reward}$ \eqref{eq:reward-bce} with respect to $\psi$
\EndFor
\EndIf
\State {{\textsc{// Policy learning}}}
\For{each timestep $t$}
\State  Collect $\state_{t+1}$ by taking $\action_t \sim \policy(\action_t | \state_t)$
\State $\mathcal{B} \leftarrow \mathcal{B}\cup \{(\state_t,\action_t,\state_{t+1},\rewmodel(\state_t))\}$
\EndFor
\For{each gradient step}
\State Sample random minibatch $\{\left(\tau_j\right)\}_{j=1}^B\sim\mathcal{B}$
\State Optimize RL objective with respect to $\phi$
\EndFor
\State Reset $\mathcal{B} \leftarrow \emptyset$ if on-policy RL algorithm is used
\EndFor
\end{algorithmic}
\end{algorithm}

\pagebreak

\section*{Experiment Setup Hyperparameters}

Our experiments use the \texttt{Walker-walk}, \texttt{Quadruped-walk}, and \texttt{Cheetah-run} environments from the DMControl suite \citep{dmcontrol}, as two of these environments (\texttt{Walker-walk} and \texttt{Quadruped-walk}) were previously used in \citet{lee2021bpref}. All hyperparameters for our method can be found in Table \ref{table:hyperparameters_ppo}, and notably, are the same as reported in \citet{lee2021bpref}. Hyperparameters that we use for the \texttt{MO-Hopper} environment are the same as those used for the \texttt{Hopper} environment in \textit{stable-baselines3}.

\begin{table}[ht]
\begin{center}
\resizebox{\columnwidth}{!}{
\begin{tabular}{l|l}
\textbf{Hyperparameter} & \textbf{Value}  \\
\hline
GAE parameter $\lambda$ & $0.9$ (Quadruped), $0.92$ (otherwise) \\ Hidden units per each layer & $256$ \\ 
Segment of length  & $50$ \\ \# of layers & $3$ \\
Learning rate  & $0.00005$ \\  Batch Size  &  $128$ (Quadruped), $64$ (otherwise)
\\  
Discount ${\bar \gamma}$ & $.99$ \\  Frequency of feedback & $32000$  \\
\# of envs per worker & $16$ (Quadruped), $32$ (otherwise)  \\ PPO clip range & $0.4$  
\\ Entropy bonus & $0.0$  \\
\# of timesteps per rollout &  $500$  \\ Maximum budget &  $2000$ (Quadruped), $1000$ (otherwise) \\
\# of feedbacks per session & $200$ (Quadruped), $100$ (otherwise) 
\end{tabular}}
\end{center}
\caption{Hyper-parameters of the \alg algorithm for \texttt{Walker-walk}, \texttt{Quadruped-walk} and \texttt{Cheetah-run} environments.}
\label{table:hyperparameters_ppo}
\end{table}

\end{document}